# ARTICLE

## Magnetomechanical Coupling in Ferronematic Phases: Influence of Spindle-Shaped Nanodopants on Liquid Crystalline Order

Karin Koch,[a] Joachim Landers,[b] Damian Günzing,[b] Hajnalka Nádasi,[c] Heiko Wende,[b] Alexey Eremin[c] and Annette M. Schmidt*[a]

Ferronematic phases, composed of liquid crystals doped with magnetic nanoparticles, exhibit unique magnetomechanical coupling effects that are of interest for responsive materials. In this study, we investigate the influence of spindle-shaped $\alpha$-Fe$_2$O$_3$ nanoparticles functionalized with a mesogen-decorated polymer brush on the phase behavior and field-induced transitions of a nematic host (5CB). Differential scanning calorimetry (DSC), refractometry, and dielectric spectroscopy reveal a non-monotonic dependence of the nematic-isotropic transition temperature on particle concentration, indicating a competition between stabilizing and destabilizing effects. The order parameter increases with increasing nanoparticle content, in contrast to non-magnetic reference systems, suggesting an alignment effect induced by the magnetic rather than the geometric anisotropy axis of the dopants. Capacitance measurements of the Fréedericksz transition show a pronounced shift in threshold fields, with a critical concentration marking a transition to enhanced magnetic responsiveness. Additional information on nanospindle diffusion correlated to the direction-dependent flow of the nematic host was inferred from comparison of Mössbauer spectroscopy and rheology data. Our findings provide insights into the interplay of magnetic and geometric anisotropy in ferronematic systems and highlight their potential for applications in tunable soft matter devices.

## 1. Introduction

Ferronematic systems, which consist of liquid crystals doped with magnetic nanoparticles, represent a promising class of soft magneto-responsive materials. [1,2,3,4,5] These hybrid systems combine the anisotropic ordering of nematic phases with the field-induced alignment properties of magnetic nanomaterials, making them attractive for applications in tuneable optical devices, adaptive lenses, and soft actuators. [6,7,8]

A key challenge in designing effective ferronematics is achieving a stable and well-defined coupling between the nematic director $\vec{n}$ and the magnetic director $\vec{m}$. Early theoretical models, such as those proposed by Brochard and de Gennes[9] and by Burylov and Raikher,[10] describe this interaction via an effective coupling constant $\Omega$, which determines whether $\vec{n}$ and $\vec{m}$ align parallel or perpendicular to each other. However, the experimental realization of strongly coupled ferronematics has remained difficult due to nanoparticle aggregation, weak magnetic anisotropy, and surface incompatibility with liquid crystals.[7]

**Ferronematic Systems: State of the Art.** Several approaches have been explored to enhance the nematic-magnetic coupling in ferronematics. Recently, important progress has been achieved through two main lines.

Surface functionalization of nanoparticles and incorporation of mesogenic ligands helped to improve the dispersibility and anchoring of the dopants within the liquid crystal host [11,12,13] and by introducing specific interactions between the particles and the liquid crystal molecules. [14] Further, optimization of the nanoparticle shape and anisotropy was suggested to maximize the response to external fields.[8]

Recent advances have provided crucial insights into important design parameters of such ferronematic dispersions. In earlier studies, we have demonstrated that surface-modified iron oxide nanoparticles significantly alter the phase behaviour and elastic properties of nematic hosts, and can be demonstrated as well for blocked as for superparamagnetic magnetic nanospheres of about 15 nm. [15,16] These works also highlight the importance of mesogenic anchoring layers, which can stabilize ferronematic phases and tune their response to external magnetic fields.[8]

Building on these findings, we explore surface compatibilized spindle-shaped $\alpha$-Fe$_2$O$_3$ nanoparticles as novel type of magnetic dopants, which offer a unique combination of shape anisotropy, surface functionalization, and magnetic responsiveness (s. Scheme 1). The dopants employ a high-aspect-ratio spindle shape, which introduces

**Motivation and Research Objectives.** In this study, we investigate the impact of spindle-shaped hematite ($\alpha$-Fe$_2$O$_3$) nanoparticles on the phase behaviour, order parameter, and field-induced transitions in ferronematic dispersions. The well-defined stabilization of particle dispersions up to moderate volume fraction of dopants is achieved by a mesogen-functionalized polymer shell to enhance nanoparticle dispersibility and anchoring interactions (s. Scheme 1). The dopants employ a high-aspect-ratio spindle shape, which introduces

a. Department for Chemistry and Biochemistry, University of Cologne, Greinstr. 4 – 6, D-507939 Köln, Germany
b. Faculty of Physics and Center for Nanointegration Duisburg-Essen (CENIDE), University of Duisburg-Essen
c. Department of Nonlinear Phenomena, Institute of Physics, Otto von Guericke University, Universitätsplatz 2, 39106 Magdeburg, Germany
* corresponding author

additional geometric anisotropy to influence nematic alignment. Systematic dielectric and birefringence measurements are combined to quantify the coupling strength and Fréedericksz transition behaviour.

Our findings provide new insights into the role of nanoparticle shape and surface chemistry in ferronematics, advancing the design of next-generation magneto-responsive liquid crystal systems.

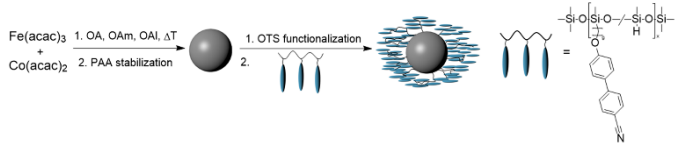

Scheme 1. Simplified scheme for the synthesis of mesogen-functional polymer-brushed nanoparticles.

## 2. Experimental Section

### 2.1 Materials

Iron(III) chloride hexahydrate (FeCl$_3$·6H$_2$O, ≥97%), sodium dihydrogen phosphate monohydrate (NaH$_2$PO$_4$·H$_2$O, ≥98%), 4-cyano-4′-pentylbiphenyl (5CB, >98%), octenyltrimethoxysilane (OTS, 95%), polymethylhydrosiloxane (PHMS, $M_n$ ≈ 1075 g mol$^{-1}$), and all solvents were obtained from commercial suppliers (Sigma-Aldrich, Merck, Acros Organics) and used without further purification. Deionized water was used for all aqueous reactions.

### 2.2 Syntheses

**Synthesis of Spindle-Shaped α-Fe$_2$O$_3$ Nanoparticles.** Spindle-shaped hematite (α-Fe$_2$O$_3$) nanoparticles were synthesized via a hydrothermal route adapted from Ozaki et al. [17] A solution of FeCl$_3$·6H$_2$O (7.015 g, 26 mmol) and NaH$_2$PO$_4$·H$_2$O (0.097 g, 0.8 mmol) was prepared in deionized water (1000 mL) and heated under reflux at 100 °C for 72 h. After cooling, the precipitate was collected by centrifugation (8500 rpm, 15 min), washed thoroughly with deionized water, and dried under vacuum.

**Surface Functionalization with Mesogenic Polymer.** To enhance dispersibility in liquid crystals, α-Fe$_2$O$_3$ nanoparticles were surface-functionalized using a mesogen-decorated polymer brush. First, OTS functionalization was performed by dispersing α-Fe$_2$O$_3$ (1 g) in ethanol (50 mL), followed by the addition of OTS (3.6 mmol). The mixture was stirred for 48 h at room temperature, and the modified particles (OTS@Fe$_2$O$_3$) were isolated by centrifugation and washing with ethanol. In a subsequent step, the mesogenic polymer poly((9-nonyloxybiphenyl carbonitril)methyl(hydro)siloxane) (9OCB-PHMS)[15] was grafted onto the OTS@Fe$_2$O$_3$ surface via a hydrosilylation reaction in dry toluene at reflux for 72 h, using hexachloroplatinic acid (0.002 eq) as a catalyst. The functionalized particles (9OCB-PHMS@Fe$_2$O$_3$) were purified by repeated precipitation in methanol and dried in vacuum.

**Preparation of Ferronematic Dispersions.** Ferronematic dispersions were prepared by dispersing 9OCB-PHMS@Fe$_2$O$_3$ nanoparticles in 5CB at various concentrations. The nanoparticles were first dispersed in dry tetrahydrofuran (THF) by ultrasonication for 30 min. The dispersion was then mixed with 5CB at isotropic conditions, followed by solvent removal under reduced pressure. The final dispersions were homogenized by vortex mixing and stored under nitrogen atmosphere.

### 2.3 Characterization Techniques

**Transmission Electron Microscopy (TEM)** Images were recorded using a LEO 912 Omega TEM (Zeiss) at 120 kV. Samples were deposited on carbon-coated copper grids from ethanol dispersions.

**Scanning Electron Microscopy (SEM)** A Zeiss Neon 40 microscope was used to analyse nanoparticle morphology. Samples were drop-cast onto silicon wafers and dried at room temperature.

**X-ray Diffraction (XRD)** Phase purity was confirmed using a Stoe-Stadi MP diffractometer (Cu Kα, λ = 1.5406 Å).

**Fourier-Transform Infrared Spectroscopy (FT-IR)** ATR-FT-IR spectra were recorded using a Shimadzu IR Affinity-1 spectrometer.

**Elemental Analysis (EA)** Carbon and nitrogen contents were measured using an Elementar Vario EL analyzer.

**Thermogravimetric Analysis (TGA)** Measurements were conducted on a PerkinElmer STA 6000 from 30 °C to 800 °C under nitrogen (heating rate: 10 K min$^{-1}$).

**Vibrating Sample Magnetometry (VSM)** Magnetization curves were obtained at room temperature using an ADE Magnetics EV7 VSM (maximum field strength: 1.2 × 10$^6$ A m$^{-1}$).

**Differential Scanning Calorimetry (DSC)** Thermal transitions were analyzed using a METTLER Toledo DSC 821e under nitrogen atmosphere. Heating/cooling cycles were performed from -50 °C to 100 °C (rate: 10 K min$^{-1}$).

**Polarized Optical Microscopy (POM)** Phase textures were observed using a Leitz Laborlux 12 POL microscope with a Linkam THMS 600 heating stage.

**Refractometry** Temperature-dependent refractive indices $n_o$ and $n_e$ were measured using a Krüss AR4 Abbé refractometer, and the order parameter $S$ was determined using the Vuks equation (see 3.2).

**Oscillating Shear Rheology (OSR)** Temperature-dependent shear viscosity of 5CB and ferronematic phases was studied using a AR-G2 rheometer (TA Instruments, New Castle, DE, USA) in plate-plate geometry (20 mm, 1 mm gap) under oscillating shear (1 Hz, 1 %) with 5 min waiting time for equilibration between the individual data points.

**Mechanical characterisation** The mechanical properties of the liquid crystal suspensions were characterized by analyzing electric and magnetic Fréedericksz transitions. Capacitance measurements were carried out using a Solartron 1260A impedance analyzer (5 kHz) and commercial planar-aligned liquid crystal cells (EHC Japan, 25 µm cell gap).

**Electric Fréedericksz Transition** Capacitance was recorded as a function of applied electric field in the range of 0.1–3.0 V at 5 kHz.

**Magnetic Fréedericksz Transition** Capacitance changes were monitored under magnetic fields ranging from 0 to 650 mT.

**Mössbauer spectroscopy (MS)** was performed in a custom-built sample holder with an integrated Peltier-element cooling module between 274 K and 315 K for a fluid sample with ca. 0.2 vol% magnetic volume fraction. The spectra were recorded in transmission geometry and constant acceleration mode using a $^{57}$Co(Rh) source.

**Data Analysis** Phase transition temperatures were obtained from differential scanning calorimetry (DSC) thermograms. The orientational order parameter $S$ was determined by fitting refractive index data to the Vuks equation (see 3.2). Threshold fields ($U_{th}$, $B_{th}$) and the splay elastic constant $K_{11}$ were extracted from capacitance data using standard Fréedericksz fitting procedures.[18]

# 3. Results and Discussion

### 3.1. Nanoparticle Characterization

The structural, morphological, and magnetic properties of the synthesized spindle-shaped α-Fe₂O₃ nanoparticles are analysed using a combination of microscopic and spectroscopic techniques. The objective is to evaluate their anisotropic shape, crystalline phase, and magnetic characteristics, which are critical for their behaviour in ferronematic dispersions.

**Morphology and Size Distribution.** Transmission electron microscopy (TEM) and scanning electron microscopy (SEM, s. Figure 1) confirm the spindle-like shape of the α-Fe₂O₃ nanoparticles. As shown in Figure 1 a, the nanoparticles exhibit an elongated morphology with an average length of (342 ± 27) nm and a width of (58.3 ± 7.5) nm, resulting in an aspect ratio of approximately 5.9. This anisotropic geometry is expected to influence their alignment within the nematic liquid crystal host.

The high aspect ratio of the particles is particularly relevant for their interaction with the nematic director, as the intrinsic easy magnetic

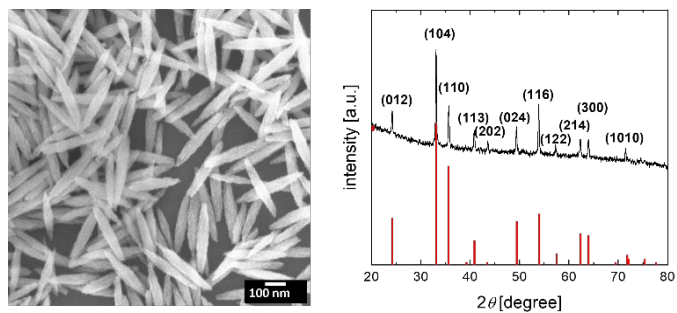

Figure 1. a) SEM image of α-Fe₂O₃ nanoparticles, b) X-ray diffractogram of α-Fe₂O₃ nanoparticles, compared with reference data.**Fehler! Textmarke nicht definiert.**

directions of the particles are along the basal plane, and thus perpendicular to their long axis. [17,19]

**Crystalline Structure.** The phase purity and crystallinity of the synthesized α-Fe₂O₃ nanoparticles are verified by X-ray diffraction (XRD). The diffraction pattern, presented in Figure 1 b, displays all peaks corresponding to the characteristic reflections of hematite (α-Fe₂O₃), in full agreement with expectations for phase-pure spinel α-Fe₂O₃ (JCPDS n°00-024-0072). No additional peaks are observed, confirming the absence of secondary phases such as maghemite (γ-Fe₂O₃) or goethite (α-FeOOH). The sharp peaks further indicates a high degree of crystallinity and the presence of single-domain particles, which is crucial for achieving distinct magnetic properties in ferronematic systems.

**Surface Functionalization and Chemical Composition** To ensure compatibility with the nematic liquid crystal host, the nanoparticles are surface-modified with a mesogen-functionalized polymer shell. The successful attachment of the polymer is confirmed by ATR-FTIR spectroscopy, as shown in Figure 2 a. The key spectral features indicative of polymer grafting include the presence of Si–H stretching vibrations at 2162 cm⁻¹, confirming the polysiloxane backbone, aromatic C≡N stretching at 2226 cm⁻¹, characteristic of the mesogenic 9OCB units, and C–H stretching vibrations (2930–2850 cm⁻¹) and biphenyl aromatic vibrations (1602–1495 cm⁻¹) that are consistent with the expected polymer composition.

Thermogravimetric analysis (TGA) further quantifies the functionalization degree, revealing a polymer content of about 14 % by weight (Table 1). Elemental analysis (EA) provides additional confirmation, with the measured carbon content matching the expected values for the grafted polymer layer.

Table 1. Mass fraction $\mu$, functionalization degree $f$, and concentration $c$ of mesogen 9OCB and individual polymer chains P[(9OCB)MS] as determined from TGA and EA measurements.

|  | $\mu$ | $f$ | $c$ |
|---|---|---|---|
|  | [m %] | [µmol·g⁻¹] | [µmol·m⁻²] |
| 9OCB | 13.7 | 920 | 73.7 |
| P[(9OCB)MS] | 5.9 | 60 | 4.7 |

**Magnetic Properties.** The magnetic response of the α-Fe₂O₃ nanoparticles is characterized using vibrating sample magnetometry (VSM) at room temperature. The magnetization curves, shown in Figure 2 b, exhibit the typical S-shape with weak ferromagnetic behaviour due to spin canting effects in hematite.

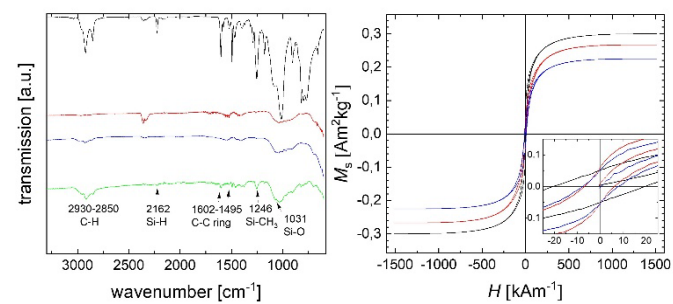

Figure 2. a) ATR-FTIR spectra and b) magnetization curves of α-Fe₂O₃ nanoparticles before and after functionalization. Black: bare α-Fe₂O₃ spindles, red: after surface priming, blue: after grafting-to of P(9OCBPHMS) to the functionalized particles, green: pure P(9OCB-PHMS).

The saturation magnetization ($M_s$) of 0.30 A·m²/kg, is lower than that of bulk α-Fe₂O₃ (0.4 A·m²/kg)[20] due to finite-size effects and surface disorder, while the coercivity ($H_c$) of 19.39 kA/m, reflects the moderate magnetic anisotropic nature of the particles. A decrease in $M_s$ after surface functionalization (to 0.22 A·m²/kg) can be attributed to the presence of the non-magnetic polymer coating. The corresponding magnetic mass fraction $\mu_{mag}$ is obtained from $\mu_{mag} = M_s/M_{s,\alpha\text{-}Fe2O3}$ ($M_{s,\alpha\text{-}Fe2O3} = 0.42$ Am²kg⁻¹)[22] for comparison. All results of the magnetic analysis are summarized in Table 2. These results confirm that the spindle-shaped α-Fe₂O₃ nanoparticles retain their intrinsic magnetic anisotropy, which plays a key role in their alignment and interaction with the nematic director in ferronematic dispersions.

Table 2. Summary of magnetic parameters from VSM measurements.

| sample | $M_s$ | $H_c$ | $M_r/M_s$ | $\mu_{mag}$ |
|---|---|---|---|---|
|  | [Am²kg⁻¹] | [kAm⁻¹] |  | [m %] |
| α-Fe₂O₃ | 0.30 | 19.39 | 0.18 | 70.8 |
| OTS@α-Fe₂O₃ | 0.28 | 6.07 | 0.19 | 69.1 |
| 9OCB-PHMS@α-Fe₂O₃ | 0.22 | 7.71 | 0.23 | 54.0 |

Thus, spindle-shaped α-Fe₂O₃ nanoparticles have been successfully synthesized. After surface functionalization with mesogenic polymer,

magnetic characterization confirms conservation of the weak ferromagnetic behaviour of the dopants.

### 3.2. Phase Behavior of Ferronematic Dispersions

The integration of spindle-shaped $\alpha$-Fe$_2$O$_3$ nanoparticles into the nematic liquid crystal host (5CB) leads to notable trends in their phase behavior. These effects are analysed using differential scanning calorimetry (DSC), polarized optical microscopy (POM), and refractometry. The primary focus is to determine how dopant concentration affects phase transition temperatures, phase stability, and the overall order of the nematic phase.

**Phase Transition Temperatures** The phase transitions of 5CB and ferronematic dispersions are investigated by DSC over a temperature range from -50 °C to 100 °C. The DSC thermograms, shown in Figure 5 a, reveal two distinct endothermic peaks. The larger transition relates to the crystalline-to-nematic transition $T_{CN}$, occurring between 20°C and 24 °C, and the smaller one to the nematic-to-isotropic transition $T_{NI}$, observed between 30°C and 36 °C. The transition temperatures extracted both from DSC and Abbé refractometry are summarized and compared in Figure 4 b. The following trends are observed:

• At low particle concentrations ($\varphi < 1 \times 10^{-4}$), both $T_{CN}$ and $T_{NI}$ show a significant increase as compared to pure 5CB, which is remarkable and indicates a stabilization of both the crystalline and the nematic phase of 5CB in the presence of the dopants. The stabilizing effect seems to be less pronounced when analysed by refractometry, though.

• Above this concentration, both $T_{CN}$ and $T_{NI}$ decrease moderately, as would be expected from a simple dilution effect caused by the magnetic dopants, compensating the stabilization. This is accompanied by a broadening of the phase transition peaks, likely due to increased heterogeneity in the domain structure.

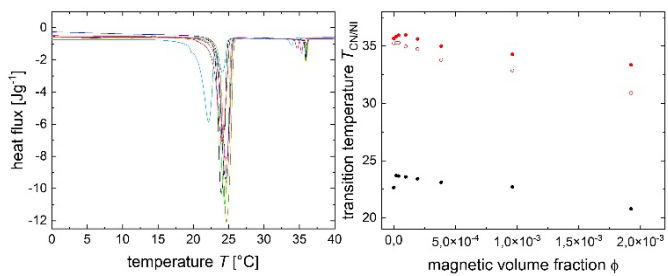

Figure 4. a) DSC thermograms of 5CB and ferronematic dispersions at different nanoparticle concentrations. 5CB (black); 9OCB-PHMS@$\alpha$-Fe$_2$O$_3$ with $\varphi$ = 1.9·10$^{-5}$ (green), $\varphi$ = 3.8 x 10$^{-5}$ (orange), $\varphi$ = 9.6·10$^{-5}$ (olive), $\varphi$ = 1.9·10$^{-4}$ (royal), $\varphi$ = 3.8·10$^{-4}$ (pink), $\varphi$ = 9.6·10$^{-4}$ (red), $\varphi$ = 1.9·10$^{-3}$ (cyan). b) (Peak) transition temperatures $T_{CN}$ (black) and $T_{NI}$ (red) as obtained by DSC (full symbols) and by Abbé refractometry (open symbols).

The transition enthalpies ($\Delta H$) are extracted from the DSC data. The crystalline-to-nematic transition enthalpy vary between 49 J g$^{-1}$ – 65 J g$^{-1}$, whereas the nematic-to-isotropic transition enthalpy ranged from 1.4 J g$^{-1}$ – 2.6 J g$^{-1}$ without a clear trend. This suggests that the nanoparticles primarily affect the phase behaviour of the nematic phase rather than of the crystalline phase.

**Optical Texture Analysis by Polarized Microscopy** The phase behavior was further investigated using polarized optical microscopy (POM), with representative textures of the ferronematic dispersions presented in Figure 3 b and c.

At low particle concentrations ($\varphi < 1 \times 10^{-4}$), the nematic domains appear well aligned, and no significant aggregation of nanoparticles is observed. In contrast, at higher concentrations, the domain structures become increasingly heterogenous in size, suggesting an impact of the dopants on the liquid crystalline order. This concentration-dependent behaviour corresponds well to the observations from DSC, where a broadened NI transition is observed

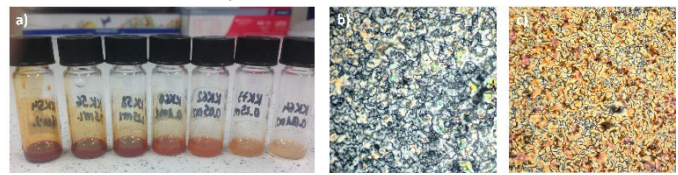

Figure 3. a) photographs and b), c) POM images of ferronematic dispersions at $\varphi$ = 9.6 x 10$^{-5}$ (b) and $\varphi$ = 9.6 x 10$^{-4}$ (c).

on the temperature scale.

**Order Parameter and Anisotropy from Refractometry** To quantify the impact of $\alpha$-Fe$_2$O$_3$ nanoparticles on the nematic order, the refractive indices ($n_o$ and $n_e$) are measured as a function of temperature using Abbé refractometry. The birefringence $\Delta n = n_e - n_o$ is used to calculate the nematic order parameter $S$ using the Vuks equation:[21]

$$S\left(\frac{\Delta\alpha}{\alpha}\right) = \frac{2(n_e^2 - n_o^2)}{n_e^2 + 2n_o^2 - 3} \qquad (1)$$

As illustrated in Figure 5 a exemplary for pure 5CB, the order parameter $S$ decreases with increasing temperature, as expected for nematic systems. An analogous behaviour is principally observed for particle-doped systems, with variations in the transition temperature and pre-transition behavior of $S$. However, the concentration-dependent analysis in Figure 5 b, comparing the order parameter of different similarly functionalised magnetic and non-magnetic nanoparticles systems at the same reduced temperature $\tau = T/T_{NI}$ = 0.985 reveals distinct differences. While non-magnetic, SiO$_2$-based nanoparticle dopants lead to a decrease of the order parameter already at low dopant fractions, for all magnetic systems (based on CoFe$_2$O$_4$,[15] Fe$_3$O$_4$,[16] and $\alpha$-Fe$_2$O$_3$ (this work)), and for the pure polymer 9OCB, an increase of $S$ with increasing doping fraction is observed. Accordingly, this is consistent with a positive impact on the nematic order arising from the magnetic moment of the dopants.

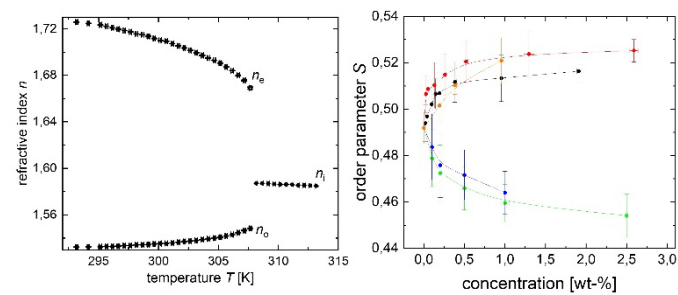

Figure 5. a) Temperature dependence of birefringence and order parameter for pure 5CB; b) order parameter at reduced temperature $\tau = T/T_{NI}$ = 0.985 as a function of nanoparticle concentration for 9OCB-PHMS@CoFe$_2$O$_4$ (black), 9OCB-PHMS@$\alpha$-Fe$_2$O$_3$ (red), 9OCB-PHMS@SiO$_2$ (25 nm, blue), 9OCB-PHMS@SiO$_2$ (300 nm, green) and pure 9OCB-PHMS (orange) in 5CB.

With this, the order parameter behaviour of ferronematic phases doped with $\alpha$-Fe$_2$O$_3$ spindles is in analogy with the observations for other magnetic nanoparticle dopants,[15,16] and deviates significantly from non-magnetic SiO$_2$ reference systems, where $S$ decreases monotonically with doping. It supports the hypothesis of a coupling between the nematic and magnetic directors, which enhances orientational order.

**Mössbauer spectroscopy.** To obtain additional information on interparticle interaction and mesogen-particle coupling effects, the dynamic behaviour of the hematite nanospindles is studied via Mössbauer spectroscopy, and their diffusion properties are correlated to the direction-dependent flow properties of the nematic host. For this, a series of Mössbauer spectra of a ferronematic dispersion with a volume fraction of $\varphi = 2 \times 10^{-3}$ is collected in a temperature range between 0 °C and 45 °C (273 K - 318 K). Exemplary spectra are shown in Figure 6 a.

All spectra show a typical sextet exhibiting the characteristic hyperfine parameters expected for hematite particles.[22] They show only minor variations over the temperature range in investigation. At the same time, the line broadening $\Delta\Gamma$ of the absorption lines becomes significantly enhanced with increasing temperature, resulting from the motion of the $^{57}$Fe-bearing particles.

The parameter $\Delta\Gamma$ can be quantitatively extracted by proper simulation of the spectra. Details on our method to reproduce spectra of hematite nanospindles and typical hyperfine parameters can be found in [22]. It has been demonstrated before that $\Delta\Gamma$ is proportional to the translational particle diffusion coefficient along γ-ray incidence direction.[23,24,25] For its interpretation, it is important to take into account that the sample is expected to be multidomain due to the absence of aligning factors, and that the spectra are collected over an extended period of time, thus showing a time average. Accordingly, the data allow the extraction of an effective translational diffusion coefficient $D_{eff}$ of the spindle-shaped dopants in the nematic host across its transition temperature, and its trend can be compared to the macroscopic sample viscosity as measured by shear rheology, which will be discussed first.

The temperature-dependent shear viscosity of pure 5CB and of a ferrofluid dispersion ($\varphi = 2 \times 10^{-3}$) measured in oscillatory experiments under low shear (Figure 6 b). The results for 5CB are in good correspondence with literature,[26] showing a steady decrease of the viscosity below the transition temperature, a local maximum of the viscosity at the phase transition, and again a decrease in the isotropic phase upon rising temperature. The absolute values are in agreement with the expectation of effective flow along the nematic director.[27] In comparison with this, the ferromagnetic sample containing nanospindles shows a similar behaviour, however, accompanied by a strong increase in the overall viscosity by a factor of 3 – 4. The transition temperature is slightly higher as for pure 5CB under these conditions, which differs from the expectations from DSC. This significant impact on the rheological behaviour is much stronger than expected when taking solely the (low) volume fraction of the solid fillers into account as an effect. On the other hand, the experimental viscosity of the ferronematic phase is similar to the viscosity of 5CB perpendicular to the nematic director (Miesowicz coefficients)[27] that are enhanced against $\eta_2$ (measured parallel to $\vec{n}$ and the flow vector $\vec{u}$) by a factor of 4.75 for $\eta_1$ (as measured parallel to $\vec{n}$ and to the flow gradient $\vec{\nabla u}$), or factor 1.86 for $\eta_3$ (as measured along the flow but perpendicular to $\vec{n}$).**Fehler! Textmarke nicht definiert.**,[27]

To compare both sets of data, the temperature-dependent diffusion coefficient extracted from the line broadening ($\Delta\Gamma$) in Mössbauer spectroscopy are displayed in Figure 6 c in comparison to the corresponding value estimated from shear viscosity. For the latter, we employ established equations for the translational diffusion of ellipsoids in fluids,[22] and, being aware of the roughness, assume that $D_{eff} = (D_{para} + 2\ D_{perp})/3$.

In parallel to the results from rheology, we observe a continuous increase of the diffusion coefficient in the nematic and supercooled nematic phase, followed by a local minimum at the nematic-isotropic transition. A good correspondence is also observed in absolute values. Overall, we see a general agreement of the effective diffusion coefficients obtained by Mössbauer spectroscopy and by shear viscosity for the ferronematic sample.

This can be interpreted as follows: The effective diffusion coefficient of an elongated particle is dominated by the diffusion along the long axis. The coincidence of the diffusion coefficients obtained by Mössbauer spectroscopy with those based on the shear viscosity of the particle dispersions are in agreement with a perpendicular alignment between the long particle axis and the nematic director of the nematic phase. In turn, at temperatures across the transition temperature (isotropic phase), the values get closer to the ones expected from the experimental viscosity of 5CB. This is despite the timescale of the technique (ca. 100 ns), which usually makes it less sensitive to long-range effects such as particle-particle interaction. The correspondence of micro- and nanoviscosity thus further supports the presence of self-assembly phenomena and ordering effects in these phases. The observation of decelerated dynamics of the hematite nanoparticles in the nematic phase is consistent with previous studies.[28,29]

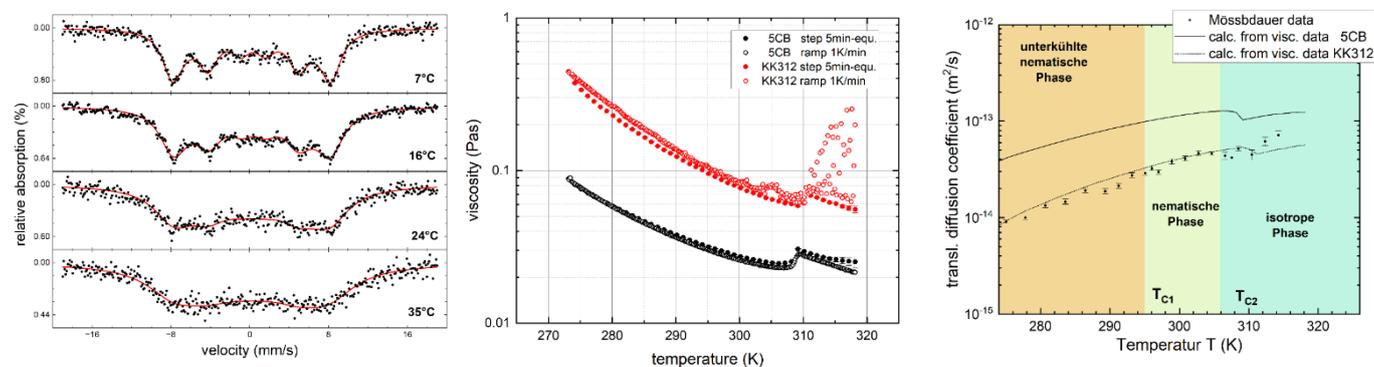

Figure 6. a) Mössbauer spectra of a 5CB-based ferronematic phase containing a volume fraction of $\varphi = 2 \times 10^{-3}$ hematite nanospindles for various temperatures between 274 K and 315 K (exemplary spectra). b) temperature-dependent shear viscosity of 5 CB (black) and a ferronematic sample with $\varphi = 2 \times 10^{-3}$; c) effective translational particle diffusion coefficient $D_{eff}$ as extracted from line broadening $\Delta\Gamma$ in Mössbauer spectroscopy (a, dots), and calculated from shear viscosity for pure 5CB (dashed) and for the ferronematic sample (solid).

# ARTICLE

**Discussion: Stabilization vs. Disruption of Nematic Order.** The observed trends in phase behavior and order parameter can be interpreted based on the interplay between geometric anchoring, magnetic interactions, and nematic elasticity. At low concentrations, the polymer-functionalized α-Fe₂O₃ nanoparticles align well with the nematic director, leading to a net stabilization effect. When the dopant fraction increases further, inter-particle interactions become more significant, leading to more heterogeneous domains in the nematic matrix, which shifts transition temperatures. The overall order parameter $S$, however, remains at high value when compared at the same reduced temperature.

Viscometry and Mössbauer spectroscopy results deliver clear indicators that even at higher volume fractions, the coupling between $\vec{n}$ and $\vec{m}$ is indicated, resulting in a preferred perpendicular configuration between the long particle axis of the magnetic dopants, and the nematic director of the system.

These findings align with previous reports on ferronematic systems,[15,16] suggesting that the balance between magnetic anisotropy and nematic elasticity is a key factor in determining the stability of ferronematic phases. This effect is to be further investigated based on electric and magnetic field-induced transitions.

### 3.3. Field-Induced Transitions

To probe the coupling between the nematic matrix and the embedded magnetic nanoparticles, electric and magnetic Fréedericksz transitions are studied using electric capacitance measurements. The field dependence of the electric capacitance reveals how the alignment of the nematic director $\vec{n}$ is influenced by both the external fields and the magnetic director $\vec{m}$. The threshold behavior and reorientation dynamics exhibit a clear dependence on particle concentration, reflecting the increasing elastic and anisotropic contributions of the dispersed phase.

**Electric Fréedericksz Transition** The electric Fréedericksz transition (FT) was studied by recording the capacitance as a function of applied voltage in glass cells with sandwich ITO electrodes and planar anchoring. The advantage of using electric field over the magnetic is that the electric field couples only to the mesogenic subphase and not to the MNPs. Figure 7 a shows the extracted effective dielectric permittivity ε as a function of applied voltage $U_{rms}$ for pure 5CB (in black). The permittivity remains nearly constant at low voltages and increases sharply beyond the threshold voltage $U_{th} \approx 750$ mV. The threshold is determined by the splay elastic constant $K_{11}$ and the dielectric anisotropy $\Delta\varepsilon = \varepsilon_\parallel - \varepsilon_\perp$ (eq. 1),[18]

$$U_{th} = \pi\sqrt{\frac{K_{11}}{\varepsilon_0 \Delta\varepsilon}} \tag{2}$$

allowing us to determine the mechanical properties of the liquid crystal, such as its splay elastic constant. These constants, along with the dielectric permittivities $\varepsilon_\perp$ and $\varepsilon_\parallel$ as extrapolated from the initial and infinite behaviour of $\varepsilon(U_{rms})$, respectively, are listed in Table 3 for different volume fractions $\varphi$.

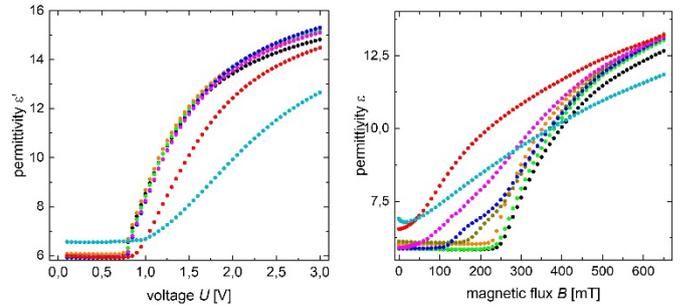

Figure 7. a) Capacitance curves for the electric Fréedericksz transition; and b) magnetic Fréedericksz transition curves for pure 5CB and for 5CB-based ferronematic phases with different particle concentrations of hematite nanospindles. 5CB (black); 9OCB-PHMS@α-Fe₂O₃ with $\varphi = 1.9 \cdot 10^{-5}$ (green), $\varphi = 3.8 \times 10^{-5}$ (orange), $\varphi = 9.6 \cdot 10^{-5}$ (olive), $\varphi = 1.9 \cdot 10^{-4}$ (royal), $\varphi = 3.8 \cdot 10^{-4}$ (pink), $\varphi = 9.6 \cdot 10^{-4}$ (red), $\varphi = 1.9 \cdot 10^{-3}$ (cyan).

The changes in the electrical response are small for ferronematic dispersions up to $\varphi \approx 3.8 \times 10^{-4}$. Beyond this concentration, the threshold voltage shifts slightly to higher values, indicating either an increase of the elastic constant or a change in $\Delta\varepsilon$ in the system (or both). A closer analysis shows that the splay elastic constant $K_{11}$ follows a similar trend, suggesting that the addition of mesogen-functionalized nanoparticles increases the elastic resistance against director reorientation. In parallel, $\varepsilon_\perp$ may also be affected by the presence of the α-Fe₂O₃ particles, particularly at higher dopant fractions. For doping levels above $4 \times 10^{-4}$, the increase both in $K_{11}$ and in $U_{th}$ becomes more pronounced, likely due to intensified particle-particle interactions and/or aggregation effects.

These results confirm that, up to moderate doping levels, the mesogenic shell promotes stronger anchoring with the nematic host, thereby increasing the energy barrier for electric field-induced reorientation.

Table 3. Threshold voltages $U_{th}$, elastic constants $K_{11}$ and dielectric constants $\varepsilon_\perp$ and $\varepsilon_\parallel$ for different particle concentrations.

| $\varphi$ | $U_{th}$ [V] | $K_{11}$ [pN] | $\varepsilon_\perp$ | $\varepsilon_\parallel$ |
|---|---|---|---|---|
| 0 | 0.750 | 5.82 | 6,05 | 17.58 |
| $1.9 \cdot 10^{-5}$ | 0.755 | 5.96 | 6.01 | 17.67 |
| $3.8 \cdot 10^{-5}$ | 0.758 | 5.89 | 6.10 | 17.52 |
| $9.6 \cdot 10^{-5}$ | 0.763 | 6.36 | 6.17 | 18.34 |
| $1.9 \cdot 10^{-4}$ | 0.755 | 6.23 | 6.05 | 18.23 |
| $3.8 \cdot 10^{-4}$ | 0.780 | 6.66 | 6.13 | 18.33 |
| $9.6 \cdot 10^{-4}$ | 0.897 | 8.93 | 6.16 | 18.53 |
| $1.9 \cdot 10^{-3}$ | 0.991 | 10.36 | 6.83 | 18.59 |

**Magnetic Fréedericksz Transition** In contrast to the electric Fréedericksz transition, which occurs at low voltages readily accessible with standard laboratory equipment, the magnetic Fréedericksz transition in conventional thermotropic liquid crystals is less known as it requires significantly stronger magnetic fields, due to the low diamagnetic anisotropy of the mesogens. To determine the magnetic threshold field $B_{th}$, capacitance measurements were performed under applied magnetic fields ranging from 0 to 650 mT. The results, summarized in Figure 6 b a, reveal a distinct dependence of the magnetic Fréedericksz transition on nanoparticle concentration. At low concentrations ($\varphi < 1 \times 10^{-4}$), the transition curve shifts toward lower magnetic fields, indicating an enhanced magnetic response. In this regime, the data can be well described by the classical Fréedericksz transition model for the magnetic case.

At higher doping levels, the field dependence exhibits an additional plateau beyond the threshold field $B_{th}$, with $B_{th}$ decreasing by nearly an order of magnitude compared to pure 5CB. In samples with $\varphi = 1 \times 10^{-3}$, the threshold field reaches about 27 mT. This reduction is attributed to coupling between the magnetic nanoparticle subphase and the liquid crystalline matrix.

As the nanoparticle concentration increases further, interparticle interactions become more prominent, leading to a more complex magneto-optic response. One approach to interpret these observations is through the Burylov–Raikher theory, which incorporates elastic and magnetic coupling effects (see below).[10] However, the empirical nature of this theory does not fully account for the microscopic anchoring behaviour at the functionally modified nanoparticle–LC interface.

**Combined Electric and Magnetic Fields** To further probe the geometries of the coupling, electric Fréedericksz transitions were measured under the simultaneous application of a constant magnetic field varied stepwise between 0 and 650 mT. Two configurations were investigated (Figure 9):

1. Parallel fields ($\vec{B} \parallel \vec{E} \perp \vec{n}$);
2. Crossed fields ($\vec{n} \parallel \vec{B} \perp \vec{E}$).

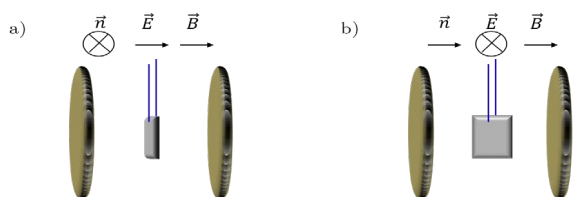

Figure 9. Schematic of experimental configurations for combined electric and magnetic field measurements of the Fréedericksz transition.

The results, shown in Figure 8 exemplarily for constant magnetic field strengths of 200 mT (parallel fields, configuration 1) and 650 mT (crossed fields, configuration 2), reveal single-step transition curves, indicating distinct trends with dopant fraction. In parallel fields, the electric response of the ferronematic phases, as compared to pure 5CB, is relatively unaffected at sub-threshold magnetic field $B$ of 200 mT (Figure 8 a). The behaviour changes for $\varphi = 3.8 \times 10^{-4}$, indicated by a sudden increase at vanishing voltage U. This trend is in accordance with the observations for the magnetic Fréedericksz transition. As shown in the double-quadratic plot (Figure 8 b), the relationship expected for a classical nematic liquid crystal ($U^2_{th} \sim B^2_{th}$) is maintained only for dopant fractions below $\varphi = 3.8 \times 10^{-4}$. For higher fractions, already small magnetic fields result in a drastic reduction of the threshold $U_{th}$ and deviation from the expected behaviour. In the parallel fields setup (configuration 1), this can be interpreted as the presence of the magnetic field pre-aligns the mesogens, thereby reducing the energy required for dielectric reorientation. In contrast, in the crossed fields setup (configuration 2), $U_{th}$ increases with increasing magnetic field strength $B$, indicating that the magnetic and electric fields compete with respect to the reorientation process. These observations confirm that the nanoparticles induce a preferential coupling between the nematic and magnetic directors, stabilizing nematic alignment parallel to the magnetic director ($\vec{n} \parallel \vec{m}$).

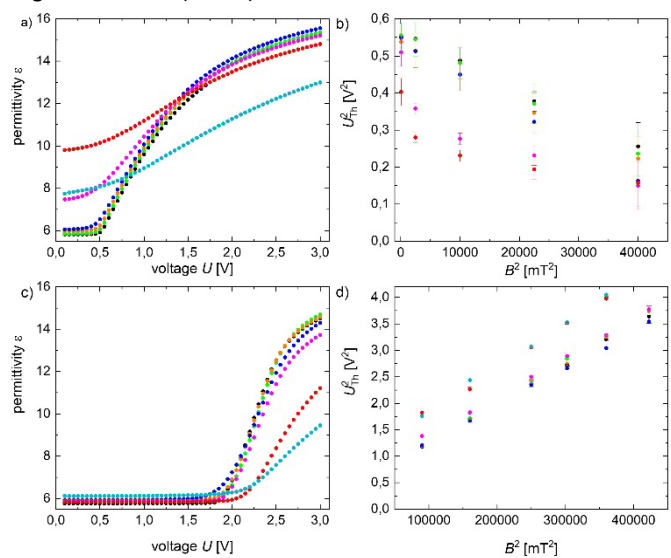

Figure 8. Electric Fréedericksz transition in constant magnetic fields (parallel vs. crossed field configurations). a) Electric Fréedericksz transition in a constant parallel magnetic field (here: at $B$ = 200 mT); b) Threshold $U_{th}^2$ of the electric Fréedericksz transition against the employed parallel magnetic field $B^2$; c) electric Fréedericksz transition in a constant perpendicular magnetic field of $B$ = 650 mT; d) Threshold $U_{th}^2$ against the employed perpendicular magnetic field $B^2$. 5CB (black); 9OCB-PHMS@α-Fe$_2$O$_3$ with $\varphi$ = 1.9·10$^{-5}$ (green), $\varphi$ = 3.8 x 10$^{-5}$ (orange), $\varphi$ = 9.6·10$^{-5}$ (olive), $\varphi$ = 1.9·10$^{-4}$ (royal), $\varphi$ = 3.8·10$^{-4}$ (pink), $\varphi$ = 9.6·10$^{-4}$ (red), $\varphi$ = 1.9·10$^{-3}$ (cyan).

**Mechanism of Magnetic-Nematic Coupling.** The observations in the magnetic and crossed fields Fréedericksz transition experiments suggest that the presence of the spindle-like α-Fe$_2$O$_3$ nanoparticles significantly modifies the nematic response to external fields in a well-defined way.

In particular, the results in the crossed field are in clear agreement with our hypothesis of a parallel alignment between the magnetic director ($\vec{m}$) and the nematic director ($\vec{n}$) of the system. At low concentrations, the nanoparticles act as individual localized alignment centers, thereby globally reducing $B_{th}$. However, if the volume fraction increases, the field response is more complex, resulting in the disappearance of the threshold behaviour and the appearance of a low-field plateau. We understand this as a competition between mesogenic anchoring and magnetic anisotropy: While the mesogen-functionalized shell strengthens particle-LC interactions, the elongated shape of the particles perpendicular to the magnetic orientation has a destabilizing effect for the nematic phase. Finally, a complex magnetic torque behaviour of the spindles in external fields and anisotropic environments**Fehler! Textmarke nicht definiert.** can be anticipated to lead to nonlinear reorientation behaviour, which still needs to be understood in detail. These findings are consistent with previous theoretical models of ferronematic coupling,[5,10] supporting the existence of a field-dependent phase transition in ferronematic dispersions and coupled magnetic and nematic directors with preferred $\vec{n} \parallel \vec{m}$ alignment.

### 3.4. Interpretation of the Ferronematic Coupling

The results presented in the previous sections highlight a parallel coupling between the nematic and magnetic directors in ferronematic dispersions doped with spindle-shaped $\alpha$-Fe$_2$O$_3$ nanoparticles. This coupling is characterized by distinct changes in phase behaviour, field-induced transitions, and critical percolation-like effects at high particle concentrations. To gain deeper insights into the nature of this interaction, we compare our findings with existing theoretical models and discuss the role of particle shape, surface anchoring, and magnetic anisotropy.

**Theoretical Background: Ferronematic Coupling Model** The behaviour of magnetic nanoparticles in liquid crystals has been described using the Burylov-Raikher theory,[10] which models the coupling between the nematic director $\vec{n}$ and the magnetic director $\vec{m}$ via a free energy density term:

$$f = \frac{1}{2}\left[K_1(\vec{\nabla}\cdot\vec{n})^2 + K_2(\vec{n}\cdot\vec{\nabla}\times\vec{n})^2 + K_3(\vec{n}\times\vec{\nabla}\times\vec{n})^2\right]$$
$$\frac{1}{2}\mu_0\Delta x(\vec{n}\cdot\vec{H})^2 - \Phi\mu_0 M_s(\vec{m}\cdot\vec{H}) - \Phi\Omega(\vec{n}\cdot\vec{m})^2 \quad (3)$$

where $\Omega$ is the coupling constant and $\vartheta$ is the angle between $\vec{n}$ and $\vec{m}$. According to this model, if $\Omega > 0$, $\vec{n}$ and $\vec{m}$ tend to align parallel (strong coupling), while if $\Omega < 0$, $\vec{n}$ and $\vec{m}$ prefer perpendicular alignment (weak or no coupling).

Qualitatively, our experimental findings suggest a positive coupling constant ($\Omega > 0$), meaning that the nematic and magnetic directors are preferentially aligned ($\vec{n} \parallel \vec{m}$). This is supported by the increase in the nematic order parameter $S$ at low concentrations, indicating an ordering effect induced by the nanoparticles (Figure 4 b). Furthermore, the decrease in $B_{th}$ for the magnetic Fréedericksz transition at low doping levels, shows enhanced magnetic susceptibility (Table 2), and the nonlinear response in combined electric and magnetic field measurements suggests a coupling between $\vec{n}$ and $\vec{m}$ (Figure 8). While this is consistent with previous studies on ferronematics, it at the same time provides new insights into the role of particle anisotropy and surface functionalization in tuning the coupling strength.

**Influence of Particle Shape and Magnetic Anisotropy** The spindle shape of $\alpha$-Fe$_2$O$_3$ nanoparticles introduces an additional level of anisotropy that influences the coupling behaviour. The preferred orientation of the magnetic moment of the $\alpha$-Fe$_2$O$_3$ spindles is perpendicular to the long axis, leading to alignment effects even at low field strength. The observed decrease in $B_{th}$ at low concentrations suggests that these particles act as local magnetic alignment centres. The long axis of the particles aligns perpendicular to the nematic director, while their magnetic easy axis lies within the basal plane.[22] This results in a preferred alignment of $\vec{m}$ along $\vec{n}$, reinforcing the coupling effect, increasing shear viscosity and the elastic constants, and reducing particle diffusivity. The combination of these factors leads to a stable ferronematic phase up to moderate doping levels.

**Role of Surface Functionalization and Anchoring Effects** The mesogen-functionalized polymer shell plays a crucial role in the stability of the ferronematic phase. The successful grafting of the mesogenic polymer to the nanoparticle surface enhances their dispersibility and compatibility with the liquid crystal host.

Key effects of the surface functionalization include an increased anchoring of the nanoparticles in the nematic matrix, leading to an increase in $U_{th}$ and $K_{11}$ in the electric Fréedericksz transition (Table 3), combined with the suppression of nanoparticle aggregation at low concentrations, which allows for effective field-induced alignment (Figure 7) and enhanced steric stabilization, preventing excessive phase separation and maintaining ferronematic properties over extended periods.

At higher concentrations, however, the influence of the dopant particles leads to more heterogeneity in the domain structure, indicated by a broadening of the NI transition and the appearance of a two-step magnetic Fréedericksz transition. Thus, the polymer shell enhances the ferronematic coupling via the particle surface, and the correlation between the shell structure and the maximum stable doping level will be of interest to investigate in future studies.

**Behaviour at High Dopant Concentrations** At a dopant level beyond ($\varphi \approx 1 \times 10^{-4}$), the domain structure of the system becomes influenced by the particle presence, characterized by increased heterogeneity disorder in the nematic phase, and by the formation of defects in POM images. These observations suggest that above a critical doping level, the system transitions from a homogeneously doped ferronematic phase to a more interphase-dominated phase. This behaviour is qualitatively similar to magnetic percolation transitions reported in other hybrid systems,[6,15,16] where localized alignment effects, probably at domain interfaces, compete with bulk nematic elasticity.

**Implications for Future Applications** The ability to tune ferronematic properties via nanoparticle shape, concentration, and surface chemistry, opens exciting possibilities for applications such as tuneable optical devices (e.g., magnetically controlled birefringence), soft actuators and artificial muscles based on field-responsive nematic materials, and adaptive lens systems with magnetically switchable focusing properties.

Further investigations could explore alternative mesogenic coatings to optimize particle-LC interactions, dynamic response studies (AC susceptibility, time-resolved birefringence), and different particle geometries and aspect ratios to fine-tune coupling strength.

## 4. Conclusions

In this study, we have investigated the magnetomechanical coupling in ferronematic phases by incorporating spindle-shaped $\alpha$-Fe$_2$O$_3$ nanoparticles into a nematic liquid crystal (5CB). The results provide new insights into the interplay between geometric anisotropy, surface functionalization, and magnetic interactions in hybrid liquid crystalline systems.

In particular, it is shown that nanoparticle anisotropy, magnetism and functionalization in combination drive ferronematic coupling. Thereby, the spindle-shaped $\alpha$-Fe$_2$O$_3$ nanoparticles exhibit a strong preference for alignment with the nematic director ($\vec{n} \parallel \vec{m}$) due to their intrinsic magnetic anisotropy. The surface functionalization with a mesogen-polymer shell enhances dispersion stability and strengthens the coupling between nanoparticles and the nematic matrix.

We show that the phase behaviour of these systems is concentration-dependent. At low doping levels ($\varphi < 1 \times 10^{-4}$), the nanoparticles enhance nematic order and phase stability, as reflected in an increase in the nematic-isotropic transition temperature ($T_{NI}$). At higher concentrations, however, nanoparticle doping leads to the emergence of higher degrees of heterogeneity and probably local defects.

Through investigation of the field-induced transitions in electric, magnetic and crossed fields, a stable and reproducible magnetic-nematic coupling is confirmed. In the electric Fréedericksz transition, the threshold voltage $U_{th}$ increases with particle concentration, confirming an increase in the elastic restoring force due to strong anchoring interactions. For the magnetically induced Fréedericksz transition, we find that at low doping levels, the threshold field $B_{th}$ decreases, indicating an enhanced response to external magnetic fields. At higher concentrations, a two-step reorientation process emerges, reflecting complex interparticle interactions.

In combined electric and magnetic fields, under either crossed or parallel fields, it is confirmed that the system exhibits a directional preference for $\vec{n} \parallel \vec{m}$, confirming a positive coupling constant ($\Omega > 0$) in agreement with the Burylov-Raikher theory. This is schematically indicated in Figure 10.

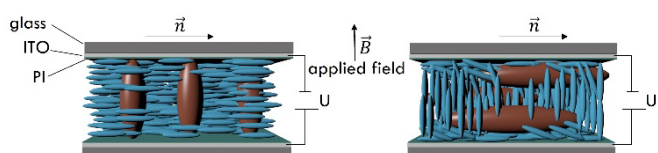

Figure 10. Cell architecture and schematic interpretation of the mechanism of enhanced magnetic response of spindle-doped ferronamtic phases.

The ability to tune the ferronematic response by adjusting particle shape, concentration, and surface chemistry opens new opportunities for applications such as tunable optical devices (e.g., magnetically controlled birefringence in adaptive lenses), soft actuators and artificial muscles and advanced display technologies based on responsive LC-based screens with magnetic control). Further investigations could explore alternative mesogenic coatings to optimize particle-LC interactions, dynamic response studies (AC susceptibility, time-resolved birefringence), and different particle geometries and aspect ratios to fine-tune coupling strength.

This comprehensive study thus demonstrates that spindle-shaped α-$Fe_2O_3$ nanoparticles act as effective magnetic dopants in liquid crystals, enhancing field responsiveness while introducing tunable coupling effects. By controlling particle anisotropy and surface functionalization, ferronematic systems can be precisely engineered for next-generation magneto-responsive materials. The findings provide a roadmap for designing next-generation magneto-responsive liquid crystalline materials.

## Author contributions

K. K.: Data curation; Formal analysis; Investigation; Methodology; Visualization; Writing – original draft

J. L.: Data curation; Formal analysis; Funding acquisition; Investigation; Methodology; Project administration; Resources; Supervision; Writing – review & editing

D. G.: Data curation; Formal analysis; Investigation: Writing – review & editing

H. N.: Data curation; Formal analysis; Investigation: Writing – review & editing

H. W.: Formal analysis; Funding acquisition; Methodology; Resources; Supervision; Validation; Writing – review & editing

A. E.: Formal analysis; Funding acquisition; Investigation; Methodology; Resources; Supervision; Validation; Writing – review & editing

A. M. S.: Conceptualization; Formal analysis; Funding acquisition; Investigation; Methodology; Project administration; Resources; Supervision; Validation; Visualization; Writing – editing  final review

## Conflicts of interest

There are no conflicts to declare.

## Data availability

Data for this article, including Origin-based primary data, are available at [to be included in final version).

## Acknowledgements

The authors are grateful for financial support from the German Science Foundation (DFG) through the priority program SPP 1681 (SCHM1747/10, STA-425/36-3 and WE2623/7-3), and follow-up projects (SCHM1747/16-1, NA 1668/1-3, and LA 5175/1-1). K. K. acknowledges support by the International Helmholtz Research School of Biophysics and Soft Matter (IHRS BioSoft). J. L. is grateful for partial support from CRC/TRR 270 (subproject B05, project-ID 405553726). We thank S. Mathur for access to XRD measurements and A. Berkessel for ATR-IR measurements.

## Notes and references